\documentstyle[12pt]{article}
\begin{document}
\def\complex{{\kern .1em {\raise .47ex \hbox
{$\scriptscriptstyle
|$}}
\kern -.4em {\rm C}}}
\def\be {\begin{equation}}
\def\ee {\end{equation}}
\def\ba{\begin{array}}
\def\ea{\end{array}}
\def\ds{\displaystyle}
\def\tr{\mbox{Tr}}
\title{Strongly Coupled Quantum and Classical Systems and
Zeno's Effect}
\author{Ph. Blanchard \\
Faculty of Physics and BiBoS, University of
Bielefeld\\
D 33615 Bielefeld, FRG
\and
A. Jadczyk \\
Institute of Theoretical Physics,
 University of Wroc{\l }aw, \\
Pl. Maxa Borna 9, PL 50 204 Wroc{\l }aw, Poland}
\maketitle{}
\vspace{2cm}
\begin{abstract}{A model interaction between a
two-state
quantum system
and a classical
switching device is analysed and shown to lead to
the
quantum Zeno
effect for large values of
the coupling costant $\kappa$. A minimal piecewise
deterministic random
process  compatible with
the  Liouville equation is described, and it is
shown that
$\kappa^{-1}$
can be interpreted
as the jump frequency of the classical device.}
\end{abstract}
\section{Introduction}
Generations of freshmen in philosophy have
discussed the
memorable
puzzles of Zeno. The
most important are about motion, \lq
Achilles\,\rq\, and the
\lq flying
arrow\,\rq\,
being the leading ones:\\
(i) Achilles cannot overtake a turtle, since
whenever he
reaches the
position from which
the turtle started the turtle has by then a new
lead.\\
(ii) If time is composed of moments, a would-be
flying arrow
must at
each moment be at
rest, and thus can never fly.\\
For quantum systems repeated frequent measurements
of
survivals can
prevent the decay of
the quantum state. This effect was formulated by
Turing 1940
and called
1977 quantum Zeno
effect by B. Misra and E.C.G. Sudarshan \cite{ms}.
In other
words:
repeated measurements
keep the state from evolving. In recent years
there has been
considerable discussion of
the quantum Zeno processes, effect and paradox.
See for
example
\cite{acc,hom,joo,fea}. Moreover it has been
claimed that
experiments
can demonstrate the effect \cite{coo,ita,pas}. In
a recent
paper
\cite{bj} we
propose a mathematically consistent model of
interaction
between
classical and quantum
systems, which provides an answer to the question
of how and
when a
quantum phenomenon
becomes real as a result of a suitable dissipative
time
evolution. With
a properly chosen
initial state the quantum probabilities are
exactly mirrored
by the
state of the classical
system and moreover the state of the quantum
subsystem
converges for
$t\rightarrow
+\infty$ to a limit which agrees with that
required by von
Neumann -
L\"uders, standard
quantum measurement projection postulate.
In our model  the quantum system is continuously
coupled to
a classical
apparatus
which will respond to its time evolution and gives
therefore
a minimal
mathematical
semantics to describe the measurement process in
Quantum
Mechanics. In
order to arrive at
Zeno's paradox one tacitly involves
Schr\"odinger's
equation. But on the
other hand we
know it holds only for an undisturbed system,
whereas in
fact we disturb
the system very
often - say, continuously - by  observations. In
Section 2
we will
briefly  describe the
model and discuss Zeno's effect in this framework.
In the
strong
coupling limit we will
estimate the distance travelled by the quantum
state $d(\hat
\rho
(0),\hat \rho (t) ),$\,
$d$ being the Bures distance in the state space.
Moreover we
give a
stochastic description
for the Zeno's model using piecewise deterministic
processes. Section 3
deals
with some concluding remarks.
\section{The Zeno's effect revisited}
Let us first very briefly describe the
mathematical
framework we will
use. For details we
refer to \cite{bj}. We consider a quantum system
$\Sigma_q$
in
interaction with a
classical system $\Sigma_c$. To the quantum system
there
corresponds a
Hilbert space ${\cal
H}_q$. In ${\cal H}_q$ we consider a family of
orthonormal
projectors
$e_i=e_i^\star=e_i^2,\; (i=1,\ldots ,n) ,\;
\sum_{i=1}^n
e_i=1\, ,$
associated to an
observable $A=\sum_{i=1}^n \lambda_i\, e_i .$ The
classical
system is
supposed to have
$m$ distinct pure states, and it is convenient to
take
$m\ge n .$ The
algebra ${\cal
A}_c$ of classical observables is in this case
${\cal A}_c =
\complex^m
.$ The set of
classical states coincides with the space of
probability
measures. Using
the notation
$X_c = \{s_0,\ldots ,s_{m-1})$, a classical state
is am
$m-$tuple
$p=(p_0,\ldots
,p_{m-1}) ,\; p_\alpha\ge 0 ,\;
\sum_{\alpha=0}^{m-1}p_\alpha =1 .$ The
state $s_0$ plays in some
cases a distinguished role and can be viewed as
the neutral
initial
state of a counter.
The algebra of observables of the total system
${\cal
A}_{tot}$ is given
by
\be {\cal A}_{tot}={\cal A}_c \otimes L({\cal
H}_q)=\complex^m\otimes
L({\cal
H}_q)=\oplus_{\alpha =0}^{m-1}L({\cal H}_q) ,\ee
and it is
convenient to
realize ${\cal A}_{tot}$
as an algebra of operators on an auxiliary Hilbert
space
${\cal
H}_{tot}={\cal H}_q\otimes
\complex^m=\oplus_{\alpha=0}^{m-1}{\cal H}_q .$
${\cal
A}_{tot}$ is then
isomorphic to the
algebra of block diagonal $m\times m$ matrices
$A=diag(a_0,a_1,\ldots
,a_{m-1})$  with
$a_\alpha\in  L({\cal H}_q).$ States on ${\cal
A}_{tot}$ are
represented
by block
diagonal matrices
\be \rho = diag (\rho_0,\rho_1,\ldots
,\rho_{m-1})\ee
where the
$\rho_\alpha $
 are positive trace class operators in
$L({\cal H}_q)$
with
$\sum_\alpha \tr(\rho_\alpha ) = 1 .$
By taking partial traces each state
$\rho $ projects on a \lq quantum state \rq
${\pi}_q(\rho )$
and a \lq classical state\rq
$\pi_c(\rho )$
given respectively by
\be \pi_q(\rho ) =\sum_\alpha \rho_\alpha,\ee
\be \pi_c(\rho ) =(\tr \rho_0,\, \tr \rho_1,\ldots
,\tr
\rho_{m-1}).\ee

The time evolution of the total system is given by
a semi
group
$\alpha^t = e^{tL}$ of completely
positive maps of ${\cal A}_{tot}$ -  preserving
hermiticity,
identity
and positivity - with $L$ of
the form \be L(A) = i[H , A ] +
{\sum}_{i = 1}^n \left ( V_i^{\star} A V_i - {1
\over 2}
\{V_i^{\star}
V_i , A\}\right )
.\label{eq:lin} \ee
The $V_i$ can be arbitrary linear operators in
$L({\cal
H}_{tot})$ such
that $\sum V_i^\star V_i
\in {\cal A}_{tot}$ and $\sum V_i^\star A V_i \in
{\cal
A}_{tot}$
whenever $A\in {\cal A}_{tot}$,
$H$ is an arbitrary block-diagonal self adjoint
operator
$H=diag(H_\alpha )$ in ${\cal H}_{tot}$
and $\{\; ,\;\}$ denotes anticommutator i.e.
\be \{A\; , B\;\} \equiv AB+BA .\ee
In order to couple the given quantum observable
$A=\sum_{i=1}^n
\lambda_i\, e_i $ to the classical
system, the $V_i$ are chosen as tensor products
$V_i=\sqrt{\kappa}
e_i\otimes\phi_i$, where
$\phi_i$ act as transformations on classical
(pure) states.
Denoting by
$\rho (t) = \alpha_t
(\rho(0))$, the time evolution of the states is
given by the
Liouville
equation
\be {\dot \rho}(t) = -i[H ,\rho (t)] +
{\sum}_{i = 1}^n \left ( V_i \rho (t) V_i^{\star}
- {1 \over
2}
\{V_i^{\star} V_i , \rho (t)\}
\right )
, \label{eq:lio}\ee
where in  general $H$ and the $V_i$ can explicitly
depend on
time.

A nice and simple example gives a model of a
continuous
measurement
where a quantum spin $1/2$
system is coupled to a 2-state classical system.
In this
situation we
consider only one orthogonal projector $e$  on the
two-dimensional
Hilbert space
${\cal H}_q=\complex^2.$ The number of classical
states is
also two. To
define the dynamics we
choose the coupling operator $V$ in the following
way:
\be
V = \sqrt{\kappa} \pmatrix{0,    & e \cr
                   e,& 0     } .\ee
The Liouville equation (\ref{eq:lio} ) for the
density
matrix $\rho =
diag(\rho_0,\rho_1)$ of the
total system reads now
\be
\ba{rl}
\ds{{\dot \rho}_0 = }\! &\ds{-i[H,\rho_0]+\kappa
(e\rho_1e-{1\over
2}\{e,\rho_0\}}),\cr\cr
\ds{{\dot \rho}_1 = }\! &\ds{-i[H,\rho_1]+\kappa
(e\rho_0e-{1\over
2}\{e,\rho_1\}).}\cr\cr
\ea
\ee
For this particularly simple  coupling the
effective quantum
state
$\hat{\rho}={\pi}_q(\rho
)=\rho_0+\rho_1$ evolves independently of the
state of the
classical
system. One can say that
here we have only transport of information from
the quantum
system to
the classical one. We have:
\be
{\dot {\hat\rho }} = -i[H,{\hat\rho}]+\kappa
(e{\hat\rho}e-{1\over
2}\{e,{\hat\rho}\}).
\ee
For the discussion of the quantum Zeno effect we
specialize:
\be
\ba{rl}
\ds{H=}\! &\ds{{\omega\over 2}{\sigma }_3,}\cr\cr
\ds{e=}\! &\ds{{1\over 2}({\sigma }_0+{\sigma
}_1),}\cr\cr
\ea
\ee
$\sigma_\mu$ being the Pauli matrices.\\
We start with the quantum system being initially
in the
eigenstate of
$\sigma_1$, and
repeatedly (with "frequency" $\kappa$ ) check if
the system
is still in
this state, each "yes"
causing a flip in the coupled classical device -
which we
can
continuously observe.

The evolution equation for ${\hat \rho}$, with the
initial
condition
${\hat\rho}(t=0)=e,$
 can be easily solved with the result:
\be
{\hat\rho}(t)={1\over
2}(\sigma_0+x(t)\sigma_1+y(t)\sigma_2),
\ee
where $x(t),y(t)$ are given by
\be
\ba{rl}
\ds{x(t)=}\! &\ds{
\exp\left(
{-\kappa t \over 4}\right)
\left(\cosh\left(
{\kappa_{\omega}t\over 4}
\right)+
{\kappa\over {\kappa_\omega}}
\sinh\left(
{\kappa_{\omega}t\over 4}
\right)\right),
}\cr\cr
\ds{y(t)=}\! &
\ds{
{4\omega\over{\kappa_\omega}}
\exp\left(
{-\kappa t \over 4}
\right)\sinh\left(
{\kappa_{\omega}t\over 4}
\right),}\cr\cr
\ea
\label{eq:xy}\ee
where $\kappa_\omega=\sqrt{\kappa^2-16\omega^2}.$
Figure 1 shows this evolution, during the time
interval
$(0,{4\pi/\omega})$ for
several different values of the dimensionless
characteristic

coefficient
\be
\alpha={\kappa\over{4\omega}}.
\ee
For $\alpha>1$ oscillations are damped completely,
and then
the distance
travelled by
the quantum state during the interaction becomes
inversely
proportional
to the square
root of
$\alpha$. The natural distance in the state space
is the
geodesic
Bures-Uhlmann distance
$d_\frown$, which is the  geodesic distance for
the
Riemannian metric
--
given in our case by $ds^2= g_{ij}dx^idx^j,$
 with $g_{ij}({\bf v})=(\delta_{ij}+v_iv_j/(1-{\bf
v}^2)).$
- cf. Refs
\cite{aj,uhl}.
 For density matrices
$v=(\sigma_0+{\bf v}\cdot {\bf \sigma})/2$ and
$w=(\sigma_0+{\bf
w}\cdot {\bf \sigma} )/2$
have (cf. Ref \cite{hub})
\be
d(v\frown w)={1\over 2}\arccos\left( {\bf
v}\cdot{\bf w}+
\root 4\of{1-{\bf v}^2}\root 4\of{1-{\bf
w}^2}\right).
 \ee
In particular, if one of the states, say $v$, is
pure, then
${\bf
v}^2=1$ and we obtain

 \be
d(v\frown w)={1\over 2}\arccos\left( {\bf
v}\cdot{\bf
w}\right).
\ee
For $v={\hat\rho} (t), w=e=(\sigma_0+\sigma_1)/2,$
as in the
Zeno model,
we obtain
\be
d({\hat\rho} (t)\frown e)={1\over 2}\arccos\left(
x(t)\right).
\label{eq:burx}\ee
Notice that $e$, being a pure state,  is on the
boundary of
the state
space, and
the $d_\frown$-distance from $e$
depends only on one of the two relevant
coordinates $x,y$  - contrary to the Frobenius
distance $\tr
((v-w)^2)$,
which
would involve both coordinates.
Assuming now $\alpha\gg 1$ and $\kappa t \gg 1$,
we get for
$x(t),y(t)$ in (\ref{eq:xy}) asymptotic formulae:
\be
\ba{rl}
\ds{x(t)}\! &\ds{\asymp 1-{2\omega^2 t\over
\kappa}+...}\cr\cr
\ds{y(t)}\! &
\ds{\asymp {\omega\over 2\kappa}+...}\cr\cr
\ea
\ee
Thus the distance  reached by state is in this
asymptotic
region given
by
\be
d \asymp \omega\sqrt{{t\over\kappa}}
\ee

The Liouville equation (\ref{eq:lio} ) describes
time
evolution of
statistical states of the
total system. The information it contains need not
be a
maximal
available one. It can be
shown  that with the equation (\ref{eq:lio}) there
is
naturally
associated a {\sl
piecewise deterministic} Markov process (cf. Ref.
\cite{dav1,dav2}) on
the set of pure states of
the total system. Knowing this process one can
answer all
kinds of
questions about time
correlations of the events, and also simulate the
random
behavior of
the
classical system coupled to a quantum one. We
refer to Ref.
\cite{sto}
for the full story, here
we will describe the particular case of interest.
\\ Let
$T_t$ be a
one-parameter semigroup of
(non-linear) transformations of rays in
$\complex^2$ given
by
\be
T(t)\phi  ={\phi (t) \over \Vert \phi (t) \Vert},
\ee
where
\be
\phi (t)=\exp\left({-iHt-{\kappa\over 2}et}\right)
\phi .
\ee
Suppose we start with the quantum
system in a pure state $\phi_0$, and  the
classical system
in a state
$s_0$. Then $\phi_0$ starts
to evolve according to the deterministic (but
non-Schr\"odinger)
evolution $T(t)\phi_0 $ until a
jump oocurs at time $t_1$. The time $t_1$ of the
jump is governed by an inhomogeneous Poisson
process with
the rate
function
$\lambda (t) = \kappa \Vert e T(t)\phi_0 \Vert^2$.
Classical
system
switches from $s_0$ to $s_1$,
while   $T(t_1)\phi_0$ jumps to
$\phi_1=eT(t_1)\phi_0$, and the process starts
again. With
the initial
state being an
eigenstate of $e$, $e\phi_0=\phi_0$, as in our
Zeno model,
and for large
values of the coupling
constant, the rate function $\lambda$ is
approximately
constant and
equal to $\kappa$. Thus
${1/\kappa}$ can be interpreted as the expected
time
interval between
the succesive jumps.
Strong coupling between the two systems, necessary
for a
manifestation
of the Zeno effect,
manifests itself with a high  frequency of jumps.

\section{Concluding Remarks}
We have shown that the mechanism leading to the
quantum Zeno
effect can
be analysed within a
model interaction between a classical and a
quantum system.
Zeno's
effect appears when a classical
device, that is able to high frequency switching
between two
alternate
states, is strongly coupled
to a quantum system prepared in an appropriate
initial
state. The
Hamiltonian evolution of the
quantum system is then slowed down, and it stops
completely
in the limit
of infinite coupling
constant. The dynamical origin of the phenomenon
is easily
understood
within our framework: the
quantum system is prepared in a initial state that
is on the
attractor
consisting of stationary
states of the dissipative part of the Liouvillian.
 The
phenomenon is
thus characteristic not
only of the particular model we have considered,
but of the
whole class
of similar models.
\\
The minimal piecewise deterministic random process
that we
have given
can be used for
computing time characteristic of the interaction,
and also
for numerical
simulations of the
phenomenon. It also shows that measuring of the
jump
frequency of the
classical apparatus can be
used for an effective estimation of the value of
the
coupling
constant.\vskip 1 cm

{\sl Acknowledgments:}
The support of Alexander von Humboldt Foundation
is
acknowledged with
thanks.

\newpage
{\bf Figure Captions}

Figure 1 : Path travelled by the quantum state
during the
time interval
$\{0,4\pi/\omega\}$, for
$\alpha=10^{-1},10^{0},10^{1},10^{2}$
respectively.
For large values of $\alpha$ quantum Zeno effect
reveals
itself: the
distance
travelled by the state is $\asymp
\sqrt{\pi\over\alpha}$
\end{document}